\documentclass[hyper]{JHEP} 

\usepackage{epsfig}

\newcommand\fverb{\setbox\pippobox=\hbox\bgroup\verb}
\newcommand\fverbdo{\egroup\medskip\noindent%
			\fbox{\unhbox\pippobox}\ }
\newcommand\fverbit{\egroup\item[\fbox{\unhbox\pippobox}]}
\newbox\pippobox
\title{Bosonic D-brane Effective Action 
 in   Linear Dilaton Background}
\author{by J. Kluso\v{n}\\
	 Department of Theoretical Physics and Astrophysics\\
                   Faculty of Science, Masaryk University\\
Kotl\'{a}\v{r}sk\'{a} 2, 611 37, Brno\\
Czech Republic\\
	E-mail: \email{klu@physics.muni.cz}}
\preprint{\hepth{0401236}}

\abstract{In this paper 
we will study  tachyon effective action for Dp-brane
in bosonic string theory in the linear dilaton
background. We obtain the tachyon effective
Lagrangian from boundary state coeficient 
of Dp-brane in the linear dilaton background
 and  compare it
with  tachyon effective Lagrangians that were
proposed in previous papers.}
\keywords{D-branes}

\def\ss{\sin \frac{\tau}{\sqrt{2}}}
\def\st{\sinh \frac{\tau}{\sqrt{2}}}
\def\st2{\sinh^2 \frac{\tau}{\sqrt{2}}}
\def\ss2{\sin^2 \frac{\tau}{\sqrt{2}}}

\begin{document}
\section{Introduction}\label{first}
Study of the tachyon condensation is
one of the central themes in the string theory research
in last few years \cite{Sen:1999mg}. 
Many  powerful methods  were developed 
in order  to describe  this process, for example
it was realised that  open
string field theory is very useful tool
 for analysis
of the tachyon condensation 
\footnote{For reviews, see 
\cite{Taylor:2003gn,Taylor:2002uv,Ohmori:2001am}.}.
In some situations one can also analyse the
tachyon condensation using 
worldsheet conformal field theory  description
\cite{Sen:1999mg,Lerda:1999um}. It was also
shown that the tachyon condensation
could be described in terms of the non-commutative
geometry \cite{Harvey:2001yn}.
One of the most remarkable results given
in last year is the fact that in some
situation the tachyon condensation could
be successfully described in terms of the
effective field theory including the tachyon
field and massless modes living on unstable
D-brane
\footnote{Some papers, where the effective
field theory descriptions of the tachyon
dynamics can be found, are
\cite{Sen:1999md,Garousi:2000tr,
Bergshoeff:2000dq,Kluson:2000iy,
Gibbons:2000hf,Kluson:2000gr,
Minahan:2000tg,Arutyunov:2000pe,Sen:2002nu,
Sen:2002an,Sen:2002qa,
Sen:2003tm,Kutasov:2003er,
Garousi:2003ai,Brax:2003rs,
Kim:2003ma,Sen:2003bc,Kluson:2003rd,
Kluson:2003sr,Smedback:2003ur,
Fotopoulos:2003yt,
Kluson:2003qk,Niarchos:2004rw}.}. 
Very nice discussion considering the
tachyon effective field theory was given recently
in \cite{Fotopoulos:2003yt}. According to
this paper  it
seems that there is no much sense to find
tachyon effective action  only since the
scale of the masses of an infinite set of
massive string modes is the same at the
tachyon mass and thus to keeping the tachyon
while integrate out all other string modes
my appear to be not well defined. 
On the other hand one may hope that some
aspects of string dynamics can be captured 
by an effective field theory action invoking
only the tachyon fields and massless modes,
where all other massive modes decouple at
a vicinity of certain conformal points. 
One such an example of an exact
conformal point  that was studied recently
is a time-dependent background which should
represent an exact boundary conformal theory
\cite{Sen:2002nu,Sen:2002an}
\begin{equation}\label{fullT}
T=f_0e^{\mu x^0}+\tilde{f}_0e^{-\mu x^0} \ ,
\end{equation}
where in the superstring case $\mu_{super}^2=\frac{1}{2}$
and in case of bosonic string theory $\mu_{bose}^2=1$. 
Its special case is the ``rolling tachyon''
background 
\begin{equation}\label{roll}
T=f_0e^{\mu x^0} \ .
\end{equation}
The disk partition function in this background
was recently studied in \cite{Larsen:2002wc}
suggesting that the corresponding potential
term should look like
\begin{equation}
V=\frac{1}{1+\frac{T^2}{2}} \ .
\end{equation}
Then it was shown in \cite{Kutasov:2003er,Niarchos:2004rw}
\footnote{For similar discussion in
case of bosonic string theory see \cite{Smedback:2003ur}.} that demanding that
a generic first-order Lagrangian should
have (\ref{fullT}) (for $\mu_{super}$) as its exact solution 
fixes its time-derivative part to be
\begin{equation}\label{kutt}
\mathcal{L}=-\frac{1}{1+\frac{T^2}{2}}\sqrt{1+\frac{T^2}{2}
-(\partial_0T)^2} \ .
\end{equation}
If we now assume as in \cite{Kutasov:2003er} that
(\ref{kutt}) has a direct Lorentz-covariant
generalisation we obtain the tachyon effective
action for unstable Dp-brane  in supersymmetric
theory
\begin{equation}\label{nkaction}
S=-\int d^{p+1}x\sqrt{-g}
\mathcal{L}  \ , 
\mathcal{L}=\frac{e^{-\Phi}}{1+\frac{T^2}{2}}
\sqrt{1+\frac{T^2}{2}+g^{\mu\nu}
\partial_{\mu}T\partial_{\nu}T}  \ . 
\end{equation}
The Lagrangian given in (\ref{nkaction}) after field redefinition
$\frac{T}{\sqrt{2}}=\sinh\frac{\tilde{T}}{\sqrt{2}}$
becomes ``tachyon DBI'' Lagrangian
\begin{equation}\label{tachyonDBI}
\mathcal{L}_{TDBI}=-e^{-\Phi}
V(\tilde{T})\sqrt{-\det(g_{\mu\nu}
+\partial_{\mu}\tilde{T}
\partial_{\nu}\tilde{T})} \ ,
V(\tilde{T})=\frac{1}
{\cosh\frac{\tilde{T}}{\sqrt{2}}} \ .
\end{equation} 
The arguments based on having (\ref{fullT}) as
an exact solution fixes only the time-derivative
dependence of the action (\ref{nkaction}). 
On the other hand it is possible that
the  covariantisation of (\ref{kutt}) to 
(\ref{nkaction}) could be rather subtle
\cite{Fotopoulos:2003yt}. In fact,
the exact description for the string partition
function evaluated on more general background
\begin{equation}
T(x^0,x^i)=f(x^0,x^i)e^{\mu x^0}
\end{equation}
should be Lorentz-covariant, but that need not
apply to its first-derivative part only. 
Exactly such a behaviour was observed in
\cite{Fotopoulos:2003yt} where 
 the string partition function in bosonic
string theory was calculated
on the tachyon profile $T=f(x^i)e^{x^0}$ 
which results into following tachyon effective
Lagrangian 
\begin{equation}\label{tselag}
\mathcal{L}=-\frac{1}{1+T}
\left[1-\frac{\ln(1+T)}{T(1+T)}
(\partial_i T)^2-\frac{s_1}{2(1+T)}
\partial^2_iT+\dots\right] \ .
\end{equation}
In supersymmetric case, where
the tachyon profile is 
\begin{equation}\label{tprof}
T=f(x^i)e^{\frac{x^0}{\sqrt{2}}} \ 
\end{equation}
we obtain tachyon effective Lagrangian
\begin{equation}
\mathcal{L}=-\frac{1}{1+\frac{T^2}{2}}
\left(1+\frac{1}{1+\frac{T^2}{2}}
\left[1-\ln\left(1+\frac{T^2}{2}\right)+\frac{1}{2}
s_1\frac{1-\frac{3}{2}T^2}{1+\frac{T^2}{2}}
\right](\partial_iT)^2+\dots\right) \ ,
\end{equation}
where $s_1$ can be changed by field redefinition.
As was shown in \cite{Fotopoulos:2003yt} this
Lagrangian differs from (\ref{nkaction}) evaluated
on the profile (\ref{tprof}).

The discussion given above implies that the form
of the tachyon effective action strongly depends
on the point in the field theory space where
this action is calculated from the string
partition function. For example, it was shown recently
in  \cite{Niarchos:2004rw} that the Lagrangian
(\ref{nkaction}) correctly reproduces  string
S-matrix tree level amplitudes near the conformal 
point $T=e^{\frac{x^0}{\sqrt{2}}}$ on condition
that  spatial momenta of  initial a final
states are small. This fact gives strong support for
the validity of (\ref{nkaction}) at least for description
of tachyon dynamics close to the 
rolling tachyon solution (\ref{roll}). On the other hand
in our recent paper
\cite{Kluson:2003sr} we have shown that there
are problems when we try to apply the action 
(\ref{nkaction}) to the case of linear dilaton
background.  The lesson from these arguments
is that we should be very careful when we try
to use tachyon effective actions in regions
of field theory space where its validity is not
established.

In this paper we will continue
the study of the tachyon effective actions in
the linear dilaton background.
 The starting point of
our analysis will be  results given in
very nice paper \cite{Karczmarek:2003xm} where
the string theory on D-branes in linear dilaton
background was studied. We use the expression
for the boundary state coefficient 
$\tilde{B}(x)$  obtained there
and following the general strategy
\cite{Tseytlin:2000mt} we will interpret
this one point function as a tachyon effective
Lagrangian evaluated on the marginal tachyon
profile $T=\pi\lambda e^{\beta t}$. 
Then in order to gain some information about
the tachyon effective action we will expand
this one point function for case of small $T$
and also for case of ``weak'' dilaton background
$V_0\ll 1$. Then we compare  resulting
effective Lagrangian  with
the tachyon effective Lagrangians  obtained in
previous papers \cite{Kutasov:2003er,Fotopoulos:2003yt}. 
We find that the tachyon effective action
obtained here has the similar form as the
tachyon effective action for D-brane in bosonic
theory (\ref{tselag}).
Then using the results given in 
\cite{Karczmarek:2003xm} we will consider
the case when the
tachyon field is still marginal however now
also depends on the spatial coordinates. 
We obtain an exact form of the string partition 
function that is manifestly Lorentz-covariant which
implies that the tachyon effective Lagrangian
evaluated on marginal tachyon profile is Lorentz-covariant
as well.  Since the form of this effective Lagrangian
is rather complicated we restrict ourselves to the case of
 weak linear
dilaton background. Once again we obtain action 
that coincides with  the action given 
in \cite{Fotopoulos:2003yt}.

This paper is organised as follows. In section
(\ref{second}) we review  results given
in \cite{Karczmarek:2003xm} that are needed for
our calculation. In section (\ref{third}) we
will consider the situation when the time-like
component of  dilaton vector is small $V_0\ll 1$ 
and we can perform
an expansion of the one point function with
respect to the small parameter $V_0$. In section
(\ref{fourth}) we generalise the calculation
given in  \cite{Karczmarek:2003xm} to the tachyon
boundary perturbation that depends on the spatial
coordinates as well. 
In conclusion (\ref{fifth}) we outline our results
and suggest possible extension of our work.
\section{Review of the tachyon condensation
on D-brane  in the linear dilaton
background}\label{second}
In this section we  briefly review the description of the
tachyon condensation in the linear dilaton background how
was presented in \cite{Karczmarek:2003xm}. 
We begin with the fact that tachyon condensate on
unstable D-brane that describes its time-dependent
decay is $T=\pi \lambda e^{\beta X^0}$ with $\beta>0$.
We will study this process in arbitrary spacelike
dilaton background 
\footnote{Our convention is $\eta_{\mu\nu}=(-1,1,\dots,p) \ ,
\mu, \nu=0,\dots, p \ , i,j=1,\dots, p$. The indices
in target spacetime are labelled with $I,J=0,\dots,D-1$.}
\begin{equation}
\Phi=V_{\mu}X^{\mu} \ , 
V_{\mu}V^{\mu}>0 \ .
\end{equation}
The weight of the worldsheet boundary interaction
operator is $\beta(\beta-V_0)$. Requiring the weight to be
equal to $1$ we obtain
\begin{equation}
V_0=\beta-\frac{1}{\beta} \ .
\end{equation}
As was shown in \cite{Karczmarek:2003xm} the time-like
part of the worldsheet CFT is the Timelike 
Boundary Liouville theory \cite{Gutperle:2003xf}
\begin{equation}\label{tlliu}
-\frac{1}{2\pi}\int_{\Sigma}
\left(\partial X^0\overline{\partial}
X^0-\frac{V_0X^0R}{4}\right)+\frac{1}{2\pi}
\int_{\partial \Sigma} (\pi \lambda  e^{\beta X^0}
+V_0 X^0 K) \ ,
\end{equation}
where $K$ is the extrinsic curvature which 
integrates to $2\pi$ around the boundary of the disc.
The action (\ref{tlliu}) is related to the
standard boundary Liouville theory with $Q=b+\frac{1}{b}$
\begin{equation}\label{slliu}
\frac{1}{2\pi}\int_{\Sigma}\left(\partial \phi
\overline{\partial}\phi+\frac{Q\phi R}{4}
\right)+\frac{1}{2\pi}\int_{\partial \Sigma}
\left(\pi\lambda e^{b\phi}+Q\phi K\right) \ ,
\end{equation}
by analytic continuation $X^0\rightarrow i\phi \ ,
\beta \rightarrow -ib \ , V_0\rightarrow -iQ$.
The boundary Liouville theory (\ref{slliu}) was
studied in
\cite{Fateev:2000ik,Teschner:2000md,
Zamolodchikov:2001ah,Ponsot:2001ng,
Schomerus:2003vv,Teschner:2003qk}.

Among many interesting results
that were calculated in  
\cite{Karczmarek:2003xm}
the most important 
  for our calculation is
the boundary state 
coefficient $\tilde{B}(x)$ that is given by
a disc worldsheet one point function
\begin{equation}\label{tildeb}
\tilde{B}(x)=e^{Vx}\int \frac{d^{D}k}{(2\pi)^D}
e^{-ikx}\left<e^{-ikX+VX}\right> \ ,
\end{equation}
where $D$ means the dimension of target spacetime. Recall
that there is  following relation between $D$ and $V_{\mu}$
in critical string theory 
\begin{equation}\label{dil}
V_{\mu}V^{\mu}=\frac{26-D}{6} \ .
\end{equation}
As in  \cite{Karczmarek:2003xm} we require
the dilaton gradient to point along the unstable
D-brane.

Now we will argue how (\ref{tildeb}) is related
to the tachyon effective Lagrangian evaluated on 
the profile $T=\pi\lambda e^{\beta x^0}$. The general definition
of the stress energy tensor is
\begin{equation}
T_{IJ}=-\frac{2}{\sqrt{-g}}
\frac{\delta S}{\delta g^{IJ}} 
 \ .
\end{equation}
If we now presume that the tachyon effective
action has
the form
\begin{equation}
S=-\int d^{p+1}x \sqrt{-g}\mathcal{L} \ 
\end{equation}
we get
\begin{equation}\label{Tgen}
T_{IJ}(x)=-g_{IJ}(x)\mathcal{L}(x)+
2\frac{\delta \mathcal{L}(x)}{\delta g^{IJ}(x)} \ .
\end{equation}
On the other hand it was shown  in \cite{Karczmarek:2003xm}
that the stress energy tensor of 
D-brane in the linear dilaton background
can be expressed in terms of the boundary
states coefficients $\tilde{A}_{\mu\nu}(x) \ ,
\tilde{B}(x)$ as 
\begin{equation}\label{Tcoef}
T_{\mu\nu}(x)=e^{-2Vx}\left(\tilde{A}
_{\mu\nu}(x)-\tilde{B}(x)\eta_{\mu\nu}\right) \ ,
\end{equation}
where the explicit form of $\tilde{A}_{\mu\nu}(x)$
is not important for us.
Comparing (\ref{Tgen}) with (\ref{Tcoef}) we
can anticipate that  the Lagrangian 
evaluated on the tachyon profile $T=\pi\lambda e^{\beta t}$
is equal to
\begin{equation}\label{bon}
\mathcal{L}(x)=
e^{-2Vx}\tilde{B}(x) \ .  
\end{equation}

The expression (\ref{tildeb}) was calculated in
\cite{Gutperle:2003xf}  with the result
\begin{equation}\label{b}
\tilde{B}(x^0)=\frac{e^{\Phi}}{\beta}\int
\frac{d ^{D-p}k}{(2\pi)^{D-p}}
e^{ikx}\tilde{\lambda}^{ik_0/\beta}
\Gamma(-ik_0/\beta)\Gamma(1+i\beta k_0) \ ,
\end{equation}
where 
\begin{equation}
\tilde{\lambda}=\frac{\pi \lambda}
{\Gamma(1+\beta^2)} \ . 
\end{equation}
Following  \cite{Karczmarek:2003xm}
we  rewrite (\ref{b}) into 
the form that will be suitable for our analysis
\begin{eqnarray}\label{bre}
\tilde{B}=\frac{e^{\Phi}}{\beta}
\int\frac{d^{D-p-1}ke^{i\sum_{i=p+1}^{D-1}k_ix^i}}
{(2\pi)^{D-p-1}}\frac{dk_0}{2\pi}
e^{ik_0x^0}\tilde{\lambda}^{ik_0/\beta}
\Gamma(-ik_0/\beta)\Gamma(1+ik_0\beta)=
\nonumber \\
=\frac{\delta(x_T)}{\beta}\int \frac{dk_0}{(2\pi)}
e^{ik_0x^0}\tilde{\lambda}^{ik_0/\beta}
\int_0^{\infty} dq e^{-q}q^{-ik_0/\beta-1}\int_0^{\infty}
 ds e^{-s}
s^{ik_0\beta}= \nonumber \\
=\delta(x_T)\int_0^{\infty} ds \int dx'^0 \int
\frac{dk_0}{(2\pi)}
e^{ik_0(x^0+\frac{1}{\beta}\ln \tilde{\lambda}-x'^0
+\beta\ln s)}e^{-e^{\beta x'^0}-s} \nonumber \\
=\delta(x_T)
\int_0^{\infty}
 ds \int dx'^0\delta(x^0+
\frac{1}{\beta}\ln \tilde{\lambda}-x'^0
+\beta\ln s)e^{-e^{\beta x'^0}-s}=
\nonumber \\
=\delta(x_T)\int_0^{\infty} ds 
\exp\left(-s-\frac{T}{\Gamma(1+\beta^2)}
s^{\beta^2}\right)
 \ , \nonumber \\
\end{eqnarray}
where we have introduced the tachyon field 
\begin{equation}\label{Tprof2}
T=\pi\lambda
e^{\beta x^0} \ ,
\end{equation} 
and where $\delta(x_T)$ means the transverse delta 
function projecting onto the Dp-brane. In what
follows we suppress this delta function.
 According to the discussion given above
 (\ref{bre}) implies that the tachyon effective
Lagrangian evaluated on the marginal profile
(\ref{Tprof2}) in the linear dilaton background
is equal to
\begin{equation}\label{bon2}
\mathcal{L}=e^{-\Phi}
\int_0^{\infty}
 ds \exp\left(-s-\frac{T}{\Gamma(1+\beta^2)}
s^{\beta^2}\right) \ .
\end{equation}
We present various checks that this Lagrangian
reduces to the more familiar one in various limits
of tachyon field theory space. 
  First of all, for $t\rightarrow -\infty$ we can write
\begin{equation}\label{lagas}
\mathcal{L}=e^{-\Phi}
\int_0^{\infty} ds e^{-s}\left(
1-\frac{ T}{\Gamma(1+\beta^2)}s^{\beta^2}\right)= \ 
e^{-\Phi}
(1-T) \ ,
\end{equation}
using 
\begin{equation}
\int_0^{\infty}ds e^{-s}=1 \ ,
 \int_0^{\infty}ds e^{-s}
s^{\beta^2}=\Gamma(\beta^2+1) \ .
\end{equation}
One can see that (\ref{lagas}) is equal 
to the bosonic version of the Lagrangian  (\ref{kutt})
given in (\ref{kutbos})  and to
 (\ref{tselag}) evaluated on time-dependent
tachyon profile in the limit of small $T$. 
Moreover, for $V_{\mu}=0$ which corresponds
to $\beta=1$, the effective Lagrangian
(\ref{bon2}) reduces to 
\begin{equation}
\mathcal{L}=\int_0^{\infty}ds e^{-s(1+T)}=
\frac{1}{1+T} \ 
\end{equation}
which is  equal to the Lagrangians
(\ref{kutbos}) and (\ref{tselag}) evaluated on
time dependent tachyon profile $T=e^t$. To gain further
information about (\ref{bon2}) we will consider
the case when  time-like component of the dilaton vector 
$V_{0}$ is  small. Let us name
this background as ``weak linear dilaton background''.
\section{Weak linear dilaton background}
\label{third}
As we said above the 
formulation weak linear dilaton background
means that   $V_0\ll 1$. Since we demand
that $\beta>0$ in order to describe tachyon
condensation where the tachyon field
rolls from $T=0$ to $T=\infty$ the 
condition of marginality   $\beta^2-\beta V_0=1$
implies  
\begin{equation}\label{betaV}
\beta=\frac{V_0+\sqrt{4+V_0^2}}{2}
\approx 1+\frac{V_0}{2} \ ,
\beta^2\approx 1+V_0 \ . 
\end{equation}
Then for small $V_0$ we get 
\begin{eqnarray}\label{LagweakV}
\mathcal{L}=e^{-\Phi}\int_0^{\infty}
ds \exp\left[-s-Ts(1+(C-1)V_0)e^{V_0\ln s}\right]=
\nonumber \\
=e^{-\Phi}\int_0^{\infty}
ds \exp\left[-s(1+T)\right]
\left(1+sV_0T\ln\frac{1}{x}+(1-C)TV_0s\right)=
\nonumber \\
=e^{-\Phi}\left(\frac{1}{1+T}+
\frac{V_0T}{(1+T)^2}
\ln (1+T)\right)=\nonumber \\
=\frac{e^{-\Phi}}{1+T}\left[1+
\frac{(\beta^2-1)T}{(1+T)}
\ln (1+T)\right] \ ,\nonumber \\
\end{eqnarray}
where we have used
\begin{eqnarray}
\Gamma(1+\beta^2)=
\Gamma(2+V_0)\approx=1-(C-1)V_0
 \ , C=0.577 \ , \nonumber \\
\int_0^{\infty}dx x e^{-ax}\ln\frac{1}{x}=
\frac{1}{a^2}\left(\ln a-1+C\right) \ . 
\end{eqnarray}
Let us say few comments about 
(\ref{LagweakV}). First of
all we must explain why we have replaced 
$V_0$ with $\beta^2-1$ using
(\ref{betaV}). We can argue as follows.
The presence of the expression $V_0T$ suggests
that the Lagrangian could  contain terms
as $\partial_{\mu}\Phi\partial_{\nu}Tg^{\mu\nu}$
that are  rather unusual and lead to
some problems. For example, the variation 
of the action containing such  terms
 with respect to $g^{\mu\nu}$
 will produce expressions
 proportional to $\partial_{\mu}\Phi\partial_{\nu}T
=\partial_{\mu}\Phi\partial_0T$. These expressions
would be presented in
 the components $T_{i0}$ of the stress energy tensor.
On the other hand we know from \cite{Karczmarek:2003xm} 
that the stress energy tensor for the rolling
tachyon solution in the linear dilaton background
is diagonal so that $T_{0i}=0$. Since the Lagrangian
(\ref{LagweakV}) is evaluated on the marginal
tachyon profile where (\ref{betaV}) holds 
we mean that it is appropriate  to
replace $V_0$ with $\beta^2-1$.
The next issue is how to interpret the expression
 $(\beta^2-1)T$.  
It can be written either $\frac{\dot{T}^2-T^2}{T}$  
or as $\ddot{T}-T$. The first possibility is non-analytic
around the point $T=0$ but together with the
factor $\ln(1+T)\sim T$ for $T\ll 1$ we get  
analytic expression.
On the other hand we mean that it is reasonable
to replace $(\beta^2-1)T$ with
$\ddot{T}-T$ in terms where the  choice
$\frac{\dot{T}^2-T^2}{T}$
 would
lead to the non-analytic behaviour around the 
point $T=0$. In fact, the importance of 
higher derivative terms in the tachyon effective
Lagrangian was discussed recently in
\cite{Fotopoulos:2003yt}. On the other hand
it is clear that the expression $(\beta^2-1)T$ is
in principle ambiguous  and that the requirement
of the analycity 
of the tachyon effective Lagrangian 
around the point $T=0$ need not be fundamental.
Moreover it is possible that
 non-analytic terms could play significant
role in some situations.
We mean that in order to resolve this issue
we should calculate the string partition 
function on more general tachyon background.
 However calculation such a partition 
function on general tachyon background
is very complicated task which is 
certainly beyond the scope of this paper.
For that reason we will 
follow the arguments given above so that 
 we write  (\ref{LagweakV})  as 
\begin{equation}\label{Tweak}
\mathcal{L}=\frac{e^{-\Phi}}{1+T}\left(1+
\frac{\ln(1+T)(\dot{T}^2-T^2)}{T(1+T)}
+\dots \right) \ . 
\end{equation}
We see that (\ref{Tweak}) has similar form as 
the Lagrangian (\ref{tselag}).  We must
also mention that in some 
sense the similarity between (\ref{Tweak}) and
(\ref{tselag})  
should not come as a big  surprise
 since both Lagrangians
are calculated from the string partition 
function even if closed string background
and hence worldsheet conformal field theories 
are different. The more interesting question
is whether there is a  relation between
(\ref{Tweak}) and the bosonic version
of (\ref{kutt}) given in \cite{Kluson:2003sr}
\footnote{For precise discussion of the
bosonic D-brane tachyon effective action,
 see also \cite{Smedback:2003ur}.}
\begin{equation}\label{kutbos}
S=- \int d^{p+1}x\frac{e^{-\Phi}}{1+T}
\sqrt{1+T-\frac{(\partial_0T)^2}{T}} \ ,
\end{equation}
where we now presume that (\ref{kutbos}) 
is valid in the nontrivial dilaton
background as well. Since for the
weak linear dilaton the factor
 $T-\frac{(\partial_0 T)^2}{T}=T(1-\beta^2)=-V_0T$ 
is small we can write 
\footnote{It is important that (\ref{kutbos}) also
implies $1-(\beta^2-1)T>1\Rightarrow T<\frac{1}{\beta^2-1}$.
On the other hand for exponential form of $T\approx e^{\beta x^0}$
there will be certainly time $x^0_*$ for which
$T(x^0_*)=\frac{1}{\beta^2-1}$. For this time
event  the Lagrangian is equal to
zero and then becomes imaginary. This fact
 again suggests
problems with the generalisation of (\ref{kutbos})
to the linear dilaton background. We can
also observe 
that the same problem arises in the supersymmetric version
(\ref{kutt}) as well.} 
\begin{equation}
\mathcal{L}=\frac{e^{-\Phi}}{1+T}
\left(1+\frac{T}{2}-\frac{(\partial_0T)^2}{2T}\right) \ 
\end{equation}
that  is clearly  different
from (\ref{Tweak}). We mean that this fact is in
agreement with the claims given in 
\cite{Fotopoulos:2003yt} that the tachyon
effective Lagrangians (\ref{kutt}),(\ref{kutbos}) do not 
follow directly from the string partition
function. We return to this issue in
the conclusion. 
\section{Spatial dependent tachyon profile}
\label{fourth}
In this section  we will consider the tachyon 
boundary interaction that also depends on
 spatial coordinates. More precisely, we
presume that the marginal interaction inserted on the
boundary of the worldsheet is 
\begin{equation}\label{Tcov}
T=\pi\lambda e^{\beta_{\mu}x^{\mu}} \ ,
-\eta^{\mu\nu}\beta_{\mu}\beta_{\nu}+
\eta^{\mu\nu}\beta_{\mu}V_{\nu}=1 \ .
\end{equation}
In this case the action for the worldsheet CFT 
has the form
\begin{equation}\label{worldaction}
S=\frac{1}{2\pi}\int_{\Sigma}
\left(\eta_{\mu\nu}\partial X^{\mu}
\overline{\partial}X^{\nu}+\frac{V_{\mu}X^{\mu}}{4}R
\right)+\frac{1}{2\pi}\int_{\partial \Sigma}
\left(\pi\lambda e^{\beta_{\mu}X^{\mu}}+V_{\mu}X^{\mu}K
\right) \ .
\end{equation}
From this explicit form of the string worldsheet action
it is clear that the exact form of the string partition
function is manifestly Lorentz covariant.
 On the other hand it is possible 
 that the expansion of the
partition function in powers of $\partial T$ will
lead to the Lagrangian where the manifest Lorentz covariance
is lost. This is an important issue that was
extensively discussed recently in 
\cite{Fotopoulos:2003yt}. One can hope that
the study the tachyon effective Lagrangian
in the linear dilaton background could
be helpful for addressing this problem. In
fact, we will see that  results given
in this section suggest possible covariant
extension of (\ref{tselag}).

First of all, using the manifest Lorentz covariance 
we can bring $\beta_{\mu}$ in (\ref{Tcov}) into the frame
where $T=\exp(\beta_{\mu}X^{\mu})=\exp(\beta_0'X^{0'})$. 
The the worldsheet action (\ref{worldaction}) is
\begin{equation}\label{worldactionT}
S=\frac{1}{2\pi}\int_{\Sigma}
\left(\eta_{\mu\nu}\partial X'^{\mu}
\overline{\partial}X'^{\nu}+\frac{V'_{\mu}X'^{\mu}}{4}R
\right)+\frac{1}{2\pi}\int_{\partial \Sigma}
\left(\pi\lambda e^{\beta'_{0}X'^{0}}+V'_{\mu}X^{'\mu}K
\right) \ 
\end{equation}
that has the same form as the worldsheet action 
given in  section (\ref{second}). Hence the 
coefficient $\tilde{B}$ will be the same 
\begin{eqnarray}
\tilde{B}(x')=e^{V_{\mu}'x^{'\mu}}\int_0^{\infty}
 ds \exp \left(-s-\frac{\pi\lambda e^{\beta' x'^0}}
{\Gamma(1+\beta'^2)}s^{\beta'^2}\right)=
\nonumber \\
=
e^{V_{\mu}x^{\mu}}\int_0^{\infty}
 ds \exp \left(-s-\frac{T}
{\Gamma(1-\eta^{\mu\nu}\beta_{\mu}\beta_{\nu})}
s^{-\eta^{\mu\nu}\beta_{\mu}\beta_{\nu}}\right) 
 \ , \nonumber \\
\end{eqnarray}
where we have introduced the tachyon field 
\begin{equation}\label{Tprofs}
T=\pi\lambda
e^{\beta_{\mu} x^{\mu}} \ 
\end{equation}
and we have also used 
 explicit Lorentz covariance  of $\tilde{B}$. 
Using the relation between $\tilde{B}$ and
$\mathcal{L}$ we obtain the tachyon effective
Lagrangian evaluated on the tachyon profile
(\ref{Tprofs}) in the form
\begin{equation}\label{Lags}
\mathcal{L}=e^{-V_{\mu}x^{\mu}}\int_0^{\infty}
 ds \exp \left(-s-\frac{T}
{\Gamma(1-\eta^{\mu\nu}\beta_{\mu}\beta_{\nu})}
s^{-\eta^{\mu\nu}\beta_{\mu}\beta_{\nu}}\right) 
\ .
\end{equation}
In order to obtain more transparent form of the Lagrangian
let us again consider the case when the dilaton 
gradient is
small. Now  the condition of marginality
implies 
\begin{equation}
-\eta^{\mu\nu}\beta_{\mu}\beta_{\nu}=
1-\eta^{\mu\nu}V_{\mu}\beta_{\nu}
=1+\epsilon  \ , \epsilon \ll 1
\end{equation}
on condition of the  weak dilaton background
$V_{\mu}\ll 1 \ , V_{\mu}V^{\mu}>0$. With this assumption
we immediately get 
\begin{eqnarray}\label{lagT2}
\mathcal{L}=e^{-V_{\mu}x^{\mu}}\int_0^{\infty}
 ds \exp \left(-s-\frac{T}
{\Gamma(1-\eta^{\mu\nu}\beta_{\mu}\beta_{\nu})}
s^{-\eta^{\mu\nu}\beta_{\mu}\beta_{\nu}}\right)= 
 \nonumber \\
=e^{-\Phi}\int ds e^{-s(1+T)}(1-T\epsilon s\ln s
+(1-C)Ts\epsilon)=
\nonumber \\
=\frac{e^{-\Phi}}{1+T}\left[
1+\frac{\epsilon T}{1+T}\ln(1+T)\right]=
\nonumber \\
=\frac{e^{-\Phi}}{1+T}\left[
1-\frac{(\eta^{\mu\nu}
\beta_{\mu}\beta_{\nu}+1) T}{1+T}\ln(1+T)]\right]
\ . 
\nonumber \\
\end{eqnarray}
 Following discussion given
in the previous section we perform the substitution
$(\eta^{\mu\nu}\beta_{\mu}\beta_{\nu}+1)T
\Rightarrow \frac{\eta^{\mu\nu}\partial_{\mu}T
\partial_{\nu}T+T^2}{T}$ in the term proportional to
$\frac{\ln(1+T)}{1+T}$ 
so that we get
\begin{equation}\label{lagc}
\mathcal{L}=\frac{e^{-\Phi}}{1+T}\left[1-\frac{\ln(1+T)}{T(1+T)}
(\eta^{\mu\nu}\partial_{\mu}T
\partial_{\nu}T+T^2)+\dots\right] \ .  
\end{equation}
The Lagrangian given above is the main result of 
this paper.  
Let us now 
presume that (\ref{lagc}) is also valid 
for general tachyon not 
obeying the  condition of marginality.
Then we see that
(\ref{lagc}) could be considered as covariant
version of (\ref{tselag}). In fact, when
 we insert
 the tachyon field  $T=f(x^i)e^{x^0}$ 
into the Lagrangian (\ref{lagc}) we get
\begin{equation}
\mathcal{L}=\frac{e^{-\Phi}}{1+T}\left[1-\frac{\ln(1+T)}{T(1+T)}
\delta^{ij}\partial_iT
\partial_jT+
\dots\right] \ . 
\end{equation}
that  
same as the first order derivative
part of the Lagrangian (\ref{tselag})
\footnote{To be precise,
the term containing the second order
derivatives of $T$ in (\ref{tselag}) arises from the renormalisation
procedure \cite{Fotopoulos:2003yt}
 that contains the beta function which is
equal to zero for the marginal tachyon profile. For
that reason it is not surprising that such term
is missing in the (\ref{lagc}) in the leading order
expansion in $\epsilon$.}.
Note also that  for  small $T$ (\ref{lagc}) has the form
\begin{eqnarray}\label{lagl}
\mathcal{L}=e^{-\Phi}\left[1-
T-\eta^{\mu\nu}\partial_{\mu}T
\partial_{\nu}T\right]
\ . 
\nonumber \\
\end{eqnarray}
We see that this Lagrangian is different from the standard
bosonic tachyon Lagrangian describing
the tachyon dynamics around the perturbative
vacuum $T=0$
\begin{equation}\label{lagpert}
\mathcal{L}=e^{-\Phi}\left[-\frac{1}{2}
\eta^{\mu\nu}\partial_{\mu}T\partial_{\nu}T
+\frac{1}{2}T^2\right] \ .
\end{equation}
Note that the same thing was recently
observed in 
\cite{Fotopoulos:2003yt} in case of
the Lagrangian (\ref{tselag}). Then it is not 
surprising that the tachyon profile 
$T=e^{\beta_{\mu}x^{\mu}}$ is 
not solution of the equation of motion
that arises from (\ref{lagl}) while is solution
of the equation of motion arising from
(\ref{lagpert})
\begin{equation}
\partial_{\mu}\left[e^{-\Phi}\eta^{\mu\nu}
\partial_{\nu}T\right]+T=0 \  
\end{equation}
for $\eta^{\mu\nu}\beta_{\mu}\beta_{\nu}-\eta^{\mu\nu}V_{\mu}
\beta_{\nu}+1=0$. 
We can explain this puzzle as follows. In 
order to derive the correct equation of motion one needs
to compute first the partition function for
the general tachyon field. In our case we
have calculated the partition function on the
exact marginal perturbation so that it is possible
that in the resulting Lagrangian 
some terms are missing which would be nonzero
for general off shell tachyon profile. 
 We also have   ambiguity
when we have replaced $\beta^2$ with either 
$\frac{\ddot{T}}{T}$ or with $\frac{\dot{T}^2}{T^2}$.
In order to obtain more general effective
Lagrangian  we should
consider arbitrary  boundary tachyon perturbation
on the string worldsheet and study this
problem using the boundary string field theory
\cite{Witten:1992qy}. 
\section{Conclusion}\label{fifth}
The main goal of this paper was to
study the tachyon effective action for
Dp-brane in the bosonic string theory in 
the linear dilaton background. We have based 
this analysis on the known form of
  the boundary state coefficient given
 in \cite{Karczmarek:2003xm} and its
relation to the tachyon effective Lagrangian
evaluated on the tachyon marginal profile. 
 As we could see in
section (\ref{second}) this Lagrangian which
is valid for general spacelike dilaton vector
$V_{\mu}$  
has rather unfamiliar form. However we have
shown that in case of vanishing dilaton field
$V_{\mu}=0$ the Lagrangian is the same as
the Lagrangians calculated for exactly
time dependent tachyon profile in
\cite{Kutasov:2003er,Fotopoulos:2003yt}.
This coincidence also holds  in the 
far past  for general $V_{\mu}$.
 Much more informations
we have got when we considered the case
when the time-like component of the dilaton
vector $V_{0}$ is small.  
Then we could introduce  small parameter
proportional to $V_0$ and perform an
expansion in the tachyon Lagrangian with
respect to it. Using the condition of the 
marginality of the tachyon profile we have got
the  tachyon effective
Lagrangian that has similar form as the
Lagrangian (\ref{tselag}). This coincidence
was much sharper when we have 
generalised the pure time dependent tachyon
condensation to the case when the tachyon profile
depends on spatial coordinates as well. 
We have argued that this can be done very easily
using manifest Lorentz-covariance of the 
boundary state coefficient 
$\tilde{B}(x)$. Then
 we have found that the resulting effective
tachyon Lagrangian  has the same  form as 
the effective tachyon Lagrangian evaluated in
\cite{Fotopoulos:2003yt}. On the other hand we
have also shown that this Lagrangian is different
from the Lagrangian (\ref{kutbos}) generalised
to the case of the linear dilaton background.
Then we have argued that this result is in 
agreement with the claim given in 
\cite{Fotopoulos:2003yt} where it was said
that (\ref{kutt}) and (\ref{kutbos}) do 
not directly follow  from the
string partition function. 
Of course this fact  does not mean that the
Lagrangian (\ref{kutt}) is not suitable
for descriptions of the tachyon condensation. 
Moreover, it was shown recently in
\cite{Niarchos:2004rw} 
 that (\ref{kutt}) gives the
excellent description of the tachyon dynamics around
the  point in the tachyon field space that
corresponds to the rolling tachyon background.
 We must also mention very interesting
paper \cite{Sen:2003zf} where it was shown
 that 
``tachyon DBI'' Lagrangian (\ref{tachyonDBI})
gives excellent description of the 
moduli space of unstable D-branes on a 
circle of critical radius. 
 On the other hand
we have shown recently in 
\cite{Kluson:2003sr}
 that the application of
(\ref{kutbos}) to the  description
of the tachyon condensation in
 the case of linear dilaton 
background leads to results which are
different from results given in
\cite{Karczmarek:2003xm}
\footnote{See also \cite{Nagami:2003mr}.}.

The extension of our calculations is obvious. 
We mean that it would be very interesting to perform the
same analysis in the supersymmetric case as well. In
order to do this we should firstly understand the
time-like form of the $N=1$ super Liouville theory 
\cite{Fukuda:2002bv,DiFrancesco:1991ud,Abdalla:hp},
following \cite{Gutperle:2003xf}.
We hope to return to this problem in future.
\\
\\
{\bf Acknowledgement}
This work was supported by the
Czech Ministry of Education under Contract No.
14310006.
\\
\\

\end{document}